\documentclass[11pt]{article}

\newcommand{\ieeeonly}[1]{}
\newcommand{\lncsonly}[1]{}
\newcommand{\articleonly}[1]{#1}
\newcommand{\acmonly}[1]{}
\newcommand{\svonly}[1]{}

\newcommand{\myparagraph}[1]{\paragraph*{#1}}

\lncsonly{\pagestyle{plain}}

\newcommand{\mycomment}[1]{}

\newcommand{\fullonly}[1]{#1} 
\newcommand{\shortonly}[1]{} 
\newcommand{\arxivonly}[1]{#1} 

\acmonly{
\settopmatter{printacmref=true}
\fancyhead{}

\def\BibTeX{{\rm B\kern-.05em{\sc i\kern-.025em b}\kern-.08emT\kern-.1667em\lower.7ex\hbox{E}\kern-.125emX}}
    
}

\usepackage{hyperref}
\usepackage{graphicx}
\usepackage{xspace}
\usepackage{amsmath}  
\usepackage[subtle,mathspacing=normal]{savetrees} 

\usepackage{multirow}
\usepackage{graphicx}

\articleonly{
\usepackage{setspace}
\setstretch{1.1}
\usepackage{fullpage} 
}

\articleonly{}
\lncsonly{}
\acmonly{}

\newcommand{\tuple}[1]{\langle #1\rangle}
\newcommand{\mean}[1]{\left[ \! \left[ #1 \right]\! \right]}

\newcommand{\intersect}{\cap}

\newcommand{\cm}{{\it CM}}
\newcommand{\om}{{\it OM}}
\newcommand{\type}{{\rm type}}
\newcommand{\stype}{{\rm sType}}
\newcommand{\scond}{{\rm sCond}}
\newcommand{\rtype}{{\rm rType}}
\newcommand{\rcond}{{\rm rCond}}
\newcommand{\con}{{\rm con}}
\newcommand{\acts}{{\rm acts}}
\newcommand{\Act}{{\it Act}}
\newcommand{\nav}{{\rm nav}}
\newcommand{\val}{{\it val}}

\newcommand{\op}{{\it op}}
\newcommand{\Rules}{{\it Rules}}
\newcommand{\spa}{{\it AU}}

\newcommand{\ind}{\hspace*{0.7em}}

\newcommand{\paths}{{\rm paths}}

\newcommand{\getpath}{{\rm path}}
\newcommand{\getsign}{{\rm sign}}
\newcommand{\getval}{{\rm val}}

\newcommand{\mspl}{{\rm MSPL}}
\newcommand{\mrpl}{{\rm MRPL}}
\newcommand{\mtpl}{{\rm MTPL}}
\newcommand{\sped}{{\rm SPED}}
\newcommand{\rped}{{\rm RPED}}
\newcommand{\mcse}{{\rm MCSE}}

\newcommand{\union}{\cup}

\newcommand{\synSim}{{\rm syn}}

\newcommand{\acSim}{\synSim_{\rm ac}}

\newcommand{\Qpol}{Q_{\rm pol}}

\newcommand{\rhomerge}{\rho_{\rm mrg}}
\newcommand{\average}{{\rm mean}}

\newcommand{\wsc}{{\rm WSC}}

\begin{document}

\acmonly{
\copyrightyear{2020} 
\acmYear{2020} 
\setcopyright{acmcopyright}
\acmConference[SACMAT '20]{25th ACM Symposium on Access Control Models and Technologies}{June 10--12, 2020}{Barcelona, Spain}
\acmBooktitle{25th ACM Symposium on Access Control Models and Technologies (SACMAT '20), June 10--12, 2020, Barcelona, Spain}
\acmPrice{15.00}
\acmDOI{10.1145/3381991.3395619}
\acmISBN{978-1-4503-7568-9/20/06}
\fancyhead{}
}

\newcommand{\thanksText}{This material is based on work supported in part by %
    NSF grants %
    CNS-1421893, 
    CCF-1414078, 
    and CCF-1954837 
    and ONR grant N00014-20-1-2751. 
    We thank Hieu Le and R. Sekar for suggesting decision tree learning as an approach to policy mining.  We thank Madison Ramos for valuable contributions to the initial stages of this work and the authors of \cite{iyer2019} for sharing their code.}
    
\title{
A Decision Tree Learning~Approach for Mining Relationship-Based Access Control Policies\ieeeonly{\thanks{\thanksText}}\lncsonly{\thanks{\thanksText}}
}

\ieeeonly{\author{
\IEEEauthorblockN{Thang Bui and Scott D. Stoller}
\IEEEauthorblockA{Department of Computer Science, Stony Brook University, USA}}}

\articleonly{\author{Thang Bui and Scott~D.~Stoller\\
  Department of Computer Science, Stony Brook University, USA}}

\lncsonly{\author{Thang Bui \and Scott~D.~Stoller}
  \institute{Department of Computer Science, Stony Brook University, USA}}

\acmonly{
\author{Thang Bui}
\affiliation{
\institution{Stony Brook University}}
\email{thang.bui@stonybrook.edu}
\author{Scott~D.~Stoller}
\affiliation{
\institution{Stony Brook University}}
\email{stoller@cs.stonybrook.edu}
}


\svonly{\author{Thang Bui \and Scott~D.~Stoller}
\institute{T. Bui \and S. D. Stoller \at Stony Brook University, USA\\ \email{stoller@cs.stonybrook.edu}}
\date{Received: date / Accepted: date}}

\newcommand{\abstracttext}{ %
Relationship-based access control (ReBAC) provides a high level of expressiveness and flexibility that promotes security and information sharing, by allowing policies to be expressed in terms of chains of relationships between entities.  ReBAC policy mining algorithms have the potential to significantly reduce the cost of migration from legacy access control systems to ReBAC, by partially automating the development of a ReBAC policy.

This paper presents new algorithms, called DTRM (Decision Tree ReBAC Miner) and DTRM$^-$, based on decision trees, for mining ReBAC policies from access control lists (ACLs) and information about entities.  Compared to state-of-the-art ReBAC mining algorithms, our algorithms are significantly faster, achieve comparable policy quality, and can mine policies in a richer language.
}

\acmonly{
\begin{abstract}
\abstracttext
\end{abstract}}


\acmonly{}


\maketitle
\ieeeonly{
\begin{abstract}
\abstracttext
\end{abstract}}
\lncsonly{
\begin{abstract}
\abstracttext
\end{abstract}}
\articleonly{
\begin{abstract}
\abstracttext
\end{abstract}}
\svonly{
\begin{abstract}
\abstracttext
\end{abstract}}


\section{Introduction}
\label{sec:intro}

In {\it relationship-based access control} (ReBAC), access control policies are expressed in terms of chains of relationships between entities.  This increases expressiveness and often allows more natural policies.  High-level access control policy models such as attribute-based access control (ABAC) and ReBAC are becoming increasingly widely adopted, as security policies become more dynamic and more complex.  ABAC is already supported by many enterprise software products\fullonly{, using a standardized ABAC language such as XACML or a vendor-specific ABAC language}.  Forms of ReBAC are supported in popular online social network systems and are being studied and adapted for use in more general software systems as well.

High-level policy models such as ReBAC allow for concise and flexible policies and promise long-term cost savings through reduced management effort.  The up-front cost of developing a ReBAC policy to replace an existing lower-level policy, such as access control lists\fullonly{ or an RBAC policy}, can be a significant barrier to adoption of ReBAC.  {\em Policy mining} algorithms have the potential to greatly reduce this cost, by automatically producing a high-level policy from existing lower-level data; vetting and tweaking it is significantly less work than creating a high-level policy from scratch.  There is\fullonly{ a substantial amount of research on role mining, surveyed in \cite{mitra2016survey,das2018}, and} a small but growing literature on ABAC policy mining\fullonly{ \cite{xu15miningABAC,xu14miningABAClogs,medvet2015,mocanu2015,sparselogs2018,iyer2018,karimi2018,cotrini2019} (surveyed in \cite{das2018})}\shortonly{, surveyed in \cite{das2018},} and ReBAC policy mining \cite{bui17mining,bui18mining,bui19mining,bui19sacmat,iyer2019}.

The ReBAC policy mining problem as defined by Bui et al. \cite{bui17mining,bui19mining} is: Given information about the attributes of all entities in the system, and the set of currently granted permissions; Find a ReBAC policy that grants the same permissions using concise, high-level rules.\fullonly{  For realistic datasets, the search space of possible policies is enormous.  In traditional ABAC languages, such as XACML, each expression involves at most one attribute dereference.  In ReBAC, an expression may contain a {\em path expression} representing a chain of attribute dereferences, and the search space grows exponentially in the path length.}


This paper proposes new ReBAC policy mining algorithms, called DTRM (Decision Tree ReBAC Miner) and DTRM$^-$, based on decision tree learning.  Decision trees are a natural basis for ReBAC policy mining because logic-based policy rules can be extracted from them much more easily than rules from neural networks,\fullonly{ Bayes classifiers,} etc.  Also, a decision tree is a compact representation of ABAC (and ReBAC) policies that supports efficient policy evaluation \cite{xengine,poltree}.  DTRM has two main phases: (1) learn an authorization policy in the form of a decision tree, using a modified version of the decision tree learning algorithm in Scikit \cite{scikitLearnDecTree}, which is an optimized version of the well-known CART algorithm \cite{cart84}, and then extract a set of candidate authorization rules from the decision tree; (2) construct the mined policy by optionally eliminating negative conditions and constraints from the candidate rules (depending on whether the target policy language is ORAL2 or ORAL2$^-$, as discussed below) and then merging and simplifying the candidate rules. 
\fullonly{We selected Scikit's algorithm because it has been used successfully in a variety of application areas, and a patch for the above modification is available for it.}


Our approach is general and could be used to mine policies in any ReBAC language.  Our implementation produces policies in an extension of ORAL (Object-oriented Relationship-based Access-control Language), a ReBAC policy language developed by Bui et al.  
ORAL \cite{bui17mining,bui18mining,bui19mining,bui19sacmat} or a similar language \cite{iyer2019} is used in much of the published work on ReBAC policy mining. 

ORAL interprets ReBAC as object-oriented ABAC: relationships are expressed using attributes that refer to other objects, and path expressions built from chains of such attributes, as in object-oriented languages such as UML and Java.  In ORAL, rules are built from {\em atomic conditions}, each of which is a condition on a single object---the subject (the entity making the access request) or resource (the entity to which access is requested)---and {\em atomic constraints}, each of which expresses a relationship between characteristics of the subject and the resource.  An example of a condition is subject.employer = LargeBank.  An example of a constraint is subject.department $\in$ resource.project.departments.

The most recent version of ORAL, introduced in \cite{bui19sacmat}, supports two additional set comparison operators.  We refer to that version as ORAL2, and we introduce ORAL2$^-$, an extension of ORAL2 with negative conditions and negative constraints.  A {\em negative} condition or constraint is the negation of an atomic condition or constraint, e.g., subject.employer $\ne$ LargeBank or subject.department $\not\in$ resource.project.departments.  We give algorithms, called DTRM and DTRM$^-$, that mine policies in ORAL2 and ORAL2$^-$, respectively.  The motivation for introducing ORAL2$^-$ is that negation is supported in some well-known ABAC languages, including XACML, and some ReBAC languages \cite{bogaerts15entity, cheng2012user, fong2011rebac-model}, and it sometimes allows more concise policies.  We also support mining ORAL2 policies, i.e., policies without negation, for two reasons.  First, some organizations may prefer policy languages without negation to reduce the chance of writing rules that grant excess permissions when new entities are added; for example, a rule with the condition subject.department $\ne$ MechEng may grant excess permissions to members of new departments, whereas a rule with the condition subject.department $\in$ \{ChemEng, ElecEng\} will not.  Second, mining of ORAL2 policies allows direct experimental comparison of our approach with FS-SEA* \cite{bui19sacmat}, a state-of-the-art ReBAC policy mining algorithm.

To demonstrate the benefits of our approach, we conducted an experimental comparison with two state-of-the-art ReBAC policy mining algorithms: FS-SEA* \cite{bui19sacmat} and Iyer et al.'s algorithm \cite{iyer2019}.  The datasets used in our experiments include four sample policies, two large case studies based on policies of real organizations \cite{decat14edoc,decat14workforce}, and several synthetic policies including the synthetic policies used in \cite{bui19sacmat}.
 
 
In summary, the main contribution of this paper is new ReBAC policy mining algorithms with two significant advantages over state-of-the-art ReBAC policy mining algorithms. (1) Our algorithms are significantly faster; specifically, they are more than $10\times$ faster than FS-SEA* on several datasets, and are several times faster than Iyer et al.'s algorithm, while achieving comparable or better quality of the mined policies. The speedup generally increases with policy size hence is expected to be even larger for the larger datasets arising in practice. (2) DTRM$^-$ mines policies in a richer language than FS-SEA* and Iyer et al.'s algorithm; specifically, the language includes set comparison operators and negation.

\section{Related Work}
\label{sec:related}

We discuss related work on ReBAC and ABAC policy mining.

\subsection{Related work on ReBAC policy mining}

Bui et al. developed several ReBAC policy mining algorithms \cite{bui17mining,bui18mining,bui19mining,bui19sacmat}, the most recent and best of which is FS-SEA* \cite{bui19sacmat}.  As shown in Section \ref{sec:evaluation-results}, our algorithms are comparably effective at discovering the desired ReBAC rules, and are significantly faster; furthermore, DTRM$^-$ can mine policies in a richer language (with negation).  Our algorithms are also simpler than FS-SEA*, which combines neural networks and a grammar-based genetic algorithm incorporating numerous heuristics
and including two stages of evolutionary search.
This is reflected in the sizes of the implementations.  There is 3 KLOC of code in common (which we copied from FS-SEA*), plus an additional 13 KLOC for FS-SEA*, compared with an additional 6 KLOC for DTRM$^-$ (our more complicated algorithm).\arxivonly{  All counts exclude blank lines and comments.}



Bui et al.'s policy mining algorithm in \cite{bui18mining}, which is a variant of the algorithm in \cite{bui19mining}, mines ReBAC policies from incomplete and noisy information about granted permissions \cite{bui18mining}.\arxivonly{  Such information is commonly available from access logs.  In particular, their algorithm identifies and removes permissions likely to be extraneous, identifies and adds permissions likely to be missing, and reports these changes to the user.}  Extending our algorithm to handle incomplete and noisy information is a direction for future work.  Decision tree pruning methods, which are designed to avoid overfitting the input data, might be suitable for this.

Iyer et al. present algorithms, based on ideas from rule mining and frequent graph-based pattern mining, for mining ReBAC authorization policies and graph transition policies \cite{iyer2019}.  Their policy mining algorithm targets a policy language that is less expressive than ORAL2$^-$, because it lacks set comparison operators and negation; furthermore, unlike ORAL2, it does not directly support Boolean attributes.   Set comparison operators are useful in practice: they are supported in XACML and used in all sample policies and case studies in \cite{bui19mining}.  Boolean attributes can be encoded in their framework, but this may require adding significant numbers of edges (connecting nodes or edges representing Boolean values to resources, since all paths referred to by a rule need to end at the resource being accessed), increasing the running time.  They experimentally compare their policy mining algorithm with Bui et al.'s greedy algorithm (however, they misinterpreted some vaguely labeled output of the tool and incorrectly reported that the greedy algorithm in \cite{bui19mining} achieved semantic similarity 0.9 for eWorkForce, while it actually achieves semantic similarity 1).  In our experiments described in Section \ref{sec:evaluation-results}, our algorithms are faster and more effective.

\subsection{Related work on ABAC Policy mining}

Xu et al. proposed the first algorithm for ABAC policy mining \cite{xu15miningABAC} and a variant of it for mining ABAC policies from logs \cite{xu14miningABAClogs}.  Medvet et al. developed the first evolutionary algorithm for ABAC policy mining \cite{medvet2015}.  Iyer et al. developed the first ABAC policy mining algorithm that can mine ABAC policies containing deny rules as well as permit rules \cite{iyer2018}.\fullonly{   Karimi et al. proposed an ABAC policy mining algorithm that uses unsupervised learning based on $k$-modes clustering \cite{karimi2018}.}  Cotrini et al. proposed a new formulation of the problem of ABAC mining from logs and a practical algorithm, called Rhapsody, to solve it \cite{sparselogs2018}.  Rhapsody is based on APRIORI-SD, a machine-learning algorithm for subgroup discovery.  Rhapsody can easily be extended to handle path expressions and therefore to support a form of ReBAC policy mining, but its running time is sensitive to the number of features and would be quite high for ReBAC mining except on small problem instances \cite{bui19sacmat}.  Cotrini et al. also developed a ``universal'' access control policy mining algorithm framework, which can be specialized to produce policy mining algorithms for a wide variety of policy languages \cite{cotrini2019}; the downside, based on their experiments, is that the resulting algorithms achieve lower policy quality than customized algorithms for specific policy languages.

A top-down approach to ABAC policy mining has been pursued, aiming to extract ABAC policies from natural language documents using natural language processing and machine learning \cite{narouei2018,alohaly2018}.



\section{Policy Language}
\label{sec:language}


Our policy language, which we call ORAL2$^-$, is Bui et al.'s ORAL2 (our name for it) \cite{bui19sacmat}, extended to allow negative conditions and constraints.  We give a brief overview of the language, and refer the reader to \cite{bui19sacmat} for details of ORAL2 and to \cite{bui19mining} for details of the original version of ORAL, which ORAL2 extends.  This overview is largely the same as in \cite{bui19sacmat}.  We include it to make this paper more self-contained, for the reader's convenience.

A {\em ReBAC policy} is a tuple $\pi=\tuple{\cm, \om, \Act, \Rules}$, where $\cm$ is a class model, $\om$ is an object model, $\Act$ is a set of actions, and $\Rules$ is a set of rules.

A {\em class model} is a set of class declarations. 
Each field has\fullonly{ a {\em type}, which is a class name or ``Boolean'', and} a {\em multiplicity}, which specifies how many values may be stored in the field and is ``one'' (also denoted ``1''), ``optional'' (also denoted ``?''), or ``many'' (also denoted ``*'', meaning any number).  Boolean fields always have multiplicity 1.  Every class implicitly contains a field ``id'' with type String and multiplicity 1.  A {\em reference type} is any class name (used as a type).\fullonly{  Like \cite{bui19sacmat}, we leave inheritance as a topic for future work.}

An {\em object model} is a set of objects whose types are consistent with the class model and with unique values in the id fields.  
Let $\type(o)$ denote the type of object $o$. 
The value of a field with multiplicity ``many'' is a set.  The value of a field with multiplicity  ``optional'' may be a single value or the placeholder $\bot$ indicating absence of a value.


A {\em path} is a sequence of field names, written with ``.'' as a separator.  
A {\em condition} is a set, interpreted as a conjunction, of atomic conditions or their negations.
An {\em atomic condition} is a tuple $\tuple{p, \op, \val}$, where $p$ is a non-empty path, $\op$ is an operator, either ``in'' or ``contains'', and $\val$ is a constant value, either an atomic value (if $\op$ is ``contains'') or a set of atomic values (if $\op$ is ``in'').  For example, an object $o$ satisfies $\tuple{{\rm dept.id}, {\rm in}, \{{\rm CompSci}\}}$ if the value obtained starting from $o$ and following (dereferencing) the dept field and then the id field equals CompSci.
In examples, conditions are usually written using mathematical notation as syntactic sugar, with ``$\in$'' for ``in'' and ``$\ni$'' for ``contains''.\fullonly{  For example, $\tuple{{\rm dept.id}, {\rm in}, \{{\rm CompSci}\}}$ is more nicely written as ${\rm dept} \in \{{\rm CompSci}\}$. Note that the path is simplified by omitting the ``id'' field since all non-Boolean paths end with ``id'' field.  Also, ``='' is used as syntactic sugar for ``in'' when the constant is a singleton set; thus, the previous example may be written as dept=CompSci.}

A {\em constraint} is a set, interpreted as a conjunction, of atomic constraints or their negations.  Informally, an atomic constraint expresses a relationship between the requesting subject and the requested resource, by relating the values of paths starting from each of them.  An {\em atomic constraint} is a tuple $\tuple{p_1, \op, p_2}$, where $p_1$ and $p_2$ are paths (possibly the empty sequence), and $\op$ is one of the following five operators: equal, in, contains, supseteq, subseteq.  
Implicitly, the first path is relative to the requesting subject, and the second path is relative to the requested resource.  The empty path represents the subject or resource itself.  For example, a subject $s$ and resource $r$ satisfy $\tuple{{\rm specialties}, {\rm contains}, {\rm topic}}$ if the set $s$.specialties contains the value $r$.topic.

In examples, constraints are written using mathematical notation as syntactic sugar, with ``$=$'' for ``equal'',``$\supseteq$'' for ``supseteq'', and ``$\subseteq$'' for ``subseteq''.

A {\em rule} is a tuple $\langle${\it subjectType}, {\it subjectCondition}, {\it resourceType}, {\it resourceCondition}, {\it constraint}, {\it actions}$\rangle$, where {\it subjectType} and {\it resourceType} are class names, {\it subjectCondition} and {\it resourceCondition} are conditions, {\it constraint} is a constraint, {\it actions} is a set of actions.  
A rule must satisfy several well-formedness requirements \cite{bui19mining}.
For a rule $\rho=\tuple{st, sc, rt, rc, c, A}$, let $\stype(\rho)=st$, $\scond(\rho)=sc$, $\rtype(\rho)=rt$, $\rcond(\rho)=rc$, $\con(\rho)=c$, and $\acts(\rho)=A$.

In example rules, paths in conditions and constraints that start from the subject and resource are prefixed with ``subject'' and ``resource'', respectively, to enhance readability. 
For example, the e-document case study \cite{bui19mining,decat14edocShort} involves a large bank whose policy contains the rule: A project member can read all sent documents regarding the project.  Using syntactic sugar, this is written as \\
$\langle\,$Employee, subject.employer = LargeBank, Document, true, subject.workOn.relatedDoc $\ni$ resource, \{read\}$\rangle$, \\
where Employee.workOn is the set of projects the employee is working on, and Project.relatedDoc is the set of sent documents related to the project.



The {\em type of a path} $p$ (relative to a specified class)
is the type of the last field in the path.  The {\em multiplicity of a path} $p$ (relative to a specified class)
is one if all fields on the path have multiplicity one, is many if any field on the path has multiplicity many, and is optional otherwise.  Given a class model, object model, object $o$, and path $p$, let $\nav(o,p)$ be the result of navigating (a.k.a. following or dereferencing) path $p$ starting from object $o$. 
The result might be no value, represented by $\bot$, an atomic value, or (if $p$ has multiplicity many) a set of values.\fullonly{
This is like the semantics of path navigation in UML's Object Constraint Language (\url{http://www.omg.org/spec/OCL/}).}

An object $o$ {\em satisfies} an atomic condition $c=\tuple{p, \op, \val}$, denoted $o\models c$, if $(\op={\rm in} \land \nav(o,p) \in \val) \lor (\op={\rm contains} \land \nav(o,p) \ni \val)$.
An object $o$ {\em satisfies} a condition $c$, denoted $o\models c$, if it satisfies each atomic condition or negated atomic condition in $c$.\fullonly{ Objects $o_1$ and $o_2$ {\em satisfy} an atomic constraint $c=\tuple{p_1, \op, p_2}$, denoted $\tuple{o_1,o_2} \models c$, if $(\op={\rm equal} \land \nav(o_1,p_1) = \nav(o_2,p_2)) \lor (\op={\rm in} \land \nav(o_1,p_1) \in \nav(o_2,p_2)) \lor (\op={\rm contains} \land \nav(o_1,p_1) \ni \nav(o_2,p_2)) \lor (\op={\rm supseteq} \land \nav(o_1,p_1) \supseteq \nav(o_2,p_2))$.
An object $o$ {\em satisfies} a constraint $c$, denoted $o\models c$, if it satisfies each atomic constraint or negated atomic constraint in $c$.}\shortonly{ Objects $o_1$ and $o_2$ {\em satisfy} an atomic constraint or constraint $c$, denoted $\tuple{o_1,o_2} \models c$, is defined in a similar way.}

An {\em SRA-tuple} is a tuple $\tuple{s, r, a}$, where the ``subject'' $s$ and ``resource'' $r$ are objects, and $a$ is an action, representing (depending on the context) authorization for $s$ to perform $a$ on $r$ or a request to perform that access.
An SRA-tuple $\tuple{s, r, a}$ {\em satisfies} a rule $\rho=\langle st, sc, rt,$ $rc, c, A\rangle$, denoted $\tuple{s, r, a} \models \rho$, if
$\type(s)=st \land s\models sc \land \type(r)=rt  \land r\models rc  \land \tuple{s,r}\models c \land a \in A$.  The {\em meaning} of a rule $\rho$, denoted $\mean{\rho}$, is the set of SRA-tuples that satisfy it.
The {\em meaning} of a ReBAC policy $\pi$, denoted $\mean{\pi}$, is the union of the meanings of its rules.


\section{Problem Definition}
\label{sec:problem}

We adopt Bui et al.'s definition of the ReBAC policy mining problem.  We present the core parts of the definition here, and refer the reader to \cite{bui19mining} for more details and discussion.

An {\em access control list (ACL) policy} is a tuple $\tuple{\cm, \om, \Act, \spa}$, where $\cm$ is a class model, $\om$ is an object model, $\Act$ is a set of actions, and $\spa\subseteq \om\times \om \times \Act$ is a set of SRA tuples representing authorizations. Conceptually, $\spa$ is the union of ACLs.  An ReBAC policy $\pi$ is {\em consistent} with an ACL policy $\langle\cm, \om,$ $\Act,$ $\spa\rangle$ if they have the same class model, object model, actions, and $\mean{\pi} = \spa$.

Among the ReBAC policies consistent with a given ACL policy $\pi_0$, the most desirable ones are those that satisfy the following two criteria.  (1) The ``id'' field should be used only when necessary, i.e., only when every ReBAC policy consistent with $\pi_0$ uses it, because uses of it make policies identity-based (like ACLs) and less general.  (2) The policy should have the best quality as measured by a given policy quality metric $\Qpol$, expressed as a function from ReBAC policies to the natural numbers, with small numbers indicating high quality.  This is natural for metrics based on policy size, which are the most common type.

The {\em ReBAC policy mining problem} is: given an ACL policy $\pi_0=\langle \cm, \om,$ $\Act, \spa\rangle$ and a policy quality metric $\Qpol$, find a set $\Rules$ of rules such that the ReBAC policy $\pi=\tuple{\cm, \om, \Act, \Rules}$ is consistent with $\pi_0$, uses the ``id'' field only when necessary, and has the best quality, according to $\Qpol$, among such policies.

The policy quality metric that our algorithm aims to optimize is {\em weighted structural complexity} (WSC), a generalization of policy size\fullonly{ first introduced for RBAC policies \cite{molloy10mining} and later extended to ReBAC} \cite{bui19mining}.  Minimizing policy size is consistent with 
usability studies showing that more concise access control policies are more manageable \cite{beckerle13formal}.  WSC is a weighted sum of the numbers of primitive elements of various kinds that appear in a rule or policy. WSC is defined bottom-up.   The $\wsc$ of an atomic condition $\tuple{p, \op, \val}$ is $|p| + |\val|$, where $|p|$ is the length of path $p$, and $|\val|$ is 1 if $\val$ is an atomic value and is the cardinality of $\val$ if $\val$ is a set.\fullonly{  The $\wsc$ of an atomic constraint $\tuple{p_1, \op, p_2}$ is $|p_1|+|p_2|$.}\shortonly{  The $\wsc$ of atomic constraints is defined similarly.}  The $\wsc$ of a negated atomic condition or constraint $c$ is 1 + $\wsc(c)$.  The WSC of a rule $\rho$, denoted $\wsc(\rho)$, is the sum of the WSCs of the atomic conditions and atomic constraints in it, plus the cardinality of the action set (more generally, it is a weighted sum of those numbers, but we take all of the weights to be 1). The WSC of a ReBAC policy $\pi$, denoted $\wsc(\pi)$, is the sum of the $\wsc$ of its rules. 


\section{Algorithm}
\label{sec:algorithm}


\subsection{Phase~1:~Learn~Decision~Tree~and Extract Rules}
\label{sec:algorithm:dt}


A {\em feature} is an atomic condition (on the subject or resource) or atomic constraint satisfying user-specified limits on lengths of paths in conditions and constraints.  We define a mapping from feature vectors to Boolean labels: given an SRA tuple $\tuple{s,r,a}$, we create a feature vector (i.e., a vector of the Boolean values of features evaluated for subject $s$ and resource $r$) and map it to true if the SRA tuple is permitted (i.e., is in $\spa$) and to false otherwise.  We represent Booleans as integers: 0 for false, and 1 for true. We train a decision tree to learn this classification (labeling) of feature vectors.

We decompose the problem based on the subject type, resource type, and action.  Specifically, we learn a separate decision tree $DT_{C_s, C_r, a}$ to classify SRA tuples with subject type $C_s$, resource type $C_r$, and action $a$.  We do this for each $\tuple{C_s, C_r, a}$ such that $\spa$ contains some SRA tuple with a subject of type $C_s$, a resource of type $C_r$, and action $a$. The inputs to $DT_{C_s, C_r, a}$ are limited to the features appropriate for subject type $C_s$ and resource type $C_r$, e.g., the path in the subject condition starts with a field in class $C_s$. The set of labeled feature vectors used to train $DT_{C_s, C_r, a}$ contains an element generated from each possible combination of a subject of type $C_s$ (in the given object model) and resource of type $C_r$.  


This decomposition by type is justified by the fact that all SRA tuples authorized by the same rule contain subjects with the same subject type and resources with the same resource type.\fullonly{  A rule can authorize SRA tuples with different actions since the last component of a rule is a set of actions.}  The first phase of our algorithm learns rules containing a single action; the second phase attempts to merge similar rules with different actions into a single rule authorizing multiple actions.

\fullonly{As an optimization, we discard a feature if it has the same truth value in all of the labeled feature vectors used to train a DT; for example, if all instances of some type $C$ in the given object model have the same value for a field $f$, then atomic conditions on field $f$ are discarded.}

We also detect sets of \textit{equivalent features}, which are features that have the same truth value in all feature vectors labeled true used to train a particular DT. For each set of equivalent features, we keep the features with the lowest WSC and discard the rest.\fullonly{  This is justified by that fact that the discarded features cannot appear in a policy with minimum WSC and consistent with $\spa$.}



Each internal node of a decision tree is labeled with a feature.  Each outgoing edge of an internal node corresponds to a possible value of the feature (true or false).  Each leaf node is labeled with an classification label (permit or deny). A feature vector is classified by testing the feature in the root node, following the edge corresponding to the value of the feature to reach a subtree, and then repeating this procedure until a leaf node is reached.

\fullonly{Figure \ref{fig:sample-dt} shows an example of a decision tree that represents a rule in an electronic medical record policy. The subject type, resource type and action are ``Physician'', ``MedicalRecord'' and ``read'', respectively.  Internal nodes and leaf nodes are represented in the figure by unfilled and filled boxes, respectively.  The rule specifies that only non-trainee physicians can read medical records which are associated to them. Formally, the rule is written as $\langle\,$Physician, subject.isTrainee = False, MedicalRecord, true, subject $\in$ resource.physician, \{read\}$\rangle$.

\begin{figure}[htb]
  \centering
\includegraphics[width=0.45\textwidth]{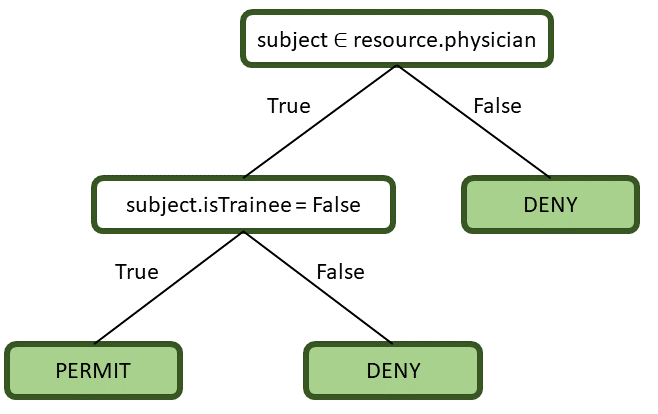}
  \caption{A sample decision tree for part of the healthcare sample policy.}
  \label{fig:sample-dt}
\end{figure}}


\subsubsection{Build Decision Trees}
\label{sec:algorithm:build-tree}

CART\fullonly{ (and other well-known decision tree building algorithms including ID3 and C4.5)} builds a decision tree by recursively partitioning feature vectors in the dataset, starting from a root node associated with the entire dataset. It chooses (as described below) a feature to test at the root node, creates a child node for each possible outcome of the test,  partitions the set of feature vectors associated with the root node among the children, based on the outcome of the test, and recursively applies this procedure to each child.  The recursion stops when all of the feature vectors associated with a node have the same classification label.


\fullonly{We use the decision tree learning algorithm in the Python library scikit-learn \cite{scikitLearnDecTree}.  It is an optimized version of CART \cite{cart84}.  We disable its pruning methods.  Pruning aims to reduce overfitting and make the decision trees generalize better.  However, a pruned tree might misclassify some feature vectors in the training data.  Pruning is therefore inappropriate for our purpose, which is to produce a policy completely consistent with the given ACL policy.  The current implementation of the algorithm in scikit-learn treats categorical features as continuous features.  For example, instead of treating a binary feature as a feature with possible values of 0 and 1, the test checks if the feature's value is less than or greater or equal than 0.5 for 0 and 1 respectively.}

To choose which feature to test at each node $n$, the algorithm applies a scoring criteria to the remaining features (i.e., features that have not been used for splitting at an ancestor of $n$) and then choosing the top-ranked feature.  The most popular scoring criteria are information gain and Gini index. For both of them, smaller values are better.
We experimented with both on some\fullonly{ sample} policies, and the generated decision trees were identical.  We adopted scikit-learn's default scoring metric, Gini index, for our experiments.

\fullonly{
\newcommand{\ig}{\mbox{\it InfoGain}}
\newcommand{\entropy}{\mbox{\it Entropy}}
\newcommand{\gini}{\mbox{\it GiniIndex}}
\newcommand{\impurity}{\mbox{\it Impurity}}
\newcommand{\lbl}{\mbox{\it label}}
\newcommand{\decis}{\mbox{\it decis}}

The Information Gain uses entropy to calculate the homogeneity of a set of feature vectors. Entropy is the measure of uncertainty of a random variable. The entropy is 0 if the sample contains only instances of the same class, and the entropy is 1 if the sample is equally divided. The information gain at node $n$ for splitting with feature $f$ is
\begin{eqnarray*}
\ig(n,f) &=& \sum_{j}\frac{|\decis(S_n,f,j)|}{|S_n|}\entropy(\decis(S_n,f,j))\\
\entropy(S) &=& -\sum_{i}\frac{|\lbl(S,i)|}{|S|} \log_{2} \frac{|\lbl(S,i)|}{|S|}
\end{eqnarray*}
where $S_n$ is the set of feature vectors associated with the current node $n$, $j$ ranges over the possible outcomes of testing feature $f$, $\decis(S,f,j)$ is the subset of $S$ containing feature vectors for which testing of feature $f$ has outcome $j$, $i$ ranges over the classification labels, and $\lbl(S,i)$ is the subset of $S$ containing feature vectors with label $i$.

The Gini Index uses impurity to measure how likely a randomly selected element would be misclassified. If all instances in the sample have the same class, the impurity will be 0. The Gini Index is calculated by subtracting the sum of squared probabilities of each class from 1. The Gini index for splitting at node $n$ with feature $f$ is
\begin{eqnarray*}
\gini(n,f) &=& \sum_{j}\frac{|\decis(S_n,f,j)|}{|S_n|}\impurity(\decis(S_n,f,j))\\
\impurity(S) &=& 1 -\sum_{i}\left(\frac{|\lbl(S,i)|}{|S|}\right)^2
\end{eqnarray*}}

When multiple features are tied for top-ranked according to the scoring criterion, scikit-learn chooses pseudorandomly among them.  We adopt a modification to the algorithm that allows specification of a secondary metric as a tie-breaker, and we use the WSC of the feature\arxivonly{ (recall that WSC of atomic conditions and atomic constraints is defined in Section \ref{sec:problem})} as the secondary metric.


\subsubsection{Extract rules}
\label{sec:algorithm:extract-rules}

We convert the decision tree into an equivalent set of rules and include them in the candidate policy.  For each distinct path through the tree from the root node to a leaf node labeled ``PERMIT'', we generate a rule containing the features associated with the internal nodes on that path; furthermore, if the path follows the False branch out of a node, then the feature associated with that node is added to the rule as a negative feature.\fullonly{
For example, for the sample decision tree in Figure \ref{fig:sample-dt}, only one rule is generated, which is the same as the input rule mentioned in Section \ref{sec:algorithm:dt}.}  Rules extracted directly from the decision tress always have non-overlapping meanings. The next phase of our algorithm can produce rules with overlapping meanings.

\subsection{Phase 2: Improve the Rules}
\label{sec:algorithm:improve}

Phase 2 has two main steps: eliminate negative features, and merge and simplify rules.

\subsubsection{Eliminate Negative Features}
\label{sec:algorithm:elim-neg}

This step is included only in DTRM, in order to mine rules without negation.  This step is omitted from DTRM$^-$.  This step eliminates each negative feature $\neg f$ in a rule $\rho$ by applying the following substeps in order until one succeeds.  A rule is {\em valid} if it covers only SRA tuples in $\spa$.

\begin{enumerate}
\item Remove $\neg f$ from $\rho$, if the resulting rule is valid.

\item Replace $\neg f$ with a feature $f'$, if the resulting rule is valid and the resulting policy (i.e., the policy with $\rho$ replaced with the resulting rule) covers all SRA tuples in $\spa$.  In particular, try this for each feature $f'$ not already used in $\rho$, in ascending order of WSC.


\item If $\neg f$ is a negative atomic condition, and path $p$ has multiplicity ``one'', then replace all of the negative atomic conditions with path $p$ with a positive atomic condition using the same path $p$ and the same operator, and with a set of constants which is the complement of the set of constants that appear in those negative atomic conditions. The complement is with respect to the set of all possible constants for that path.  Note that this step always succeeds when it is applicable, i.e., the resulting rule is always valid, and the resulting policy always covers all SRA tuples in $\spa$.\fullonly{  Generalizing this step to apply when $p$ has multiplicity ``many'' would require either replacing $\rho$ with multiple rules, or extending the policy language to allow atomic conditions containing operators (such as $\supseteq$) for which both arguments have multiplicity ``many''.}



\item If $\neg f$ is a subject atomic condition, remove all subject atomic conditions (positive or negative) in $\rho$, and add the condition ``subject.id $\in C$'', where $C$ is the set of ids of subjects that appear in SRA tuples covered by $\rho$.  An analogous step applies if $\neg f$ is a resource atomic condition.  Note that this step always succeeds when it is applicable.


\item Replace $\neg f$ with a set of features, if the resulting rule is valid and the resulting policy covers all SRA tuples in $\spa$.  In particular, try this for all sets containing two more features not already used in $\rho$, in ascending order of WSC of the set (which is the sum of the WSCs of the features in it).  Note that this step can be reached only if $f$ is a constraint.  In the experiments described in Section \ref{sec:evaluation-results}, this step is never reached, i.e., one of the previous steps always succeeds.
\end{enumerate}


\subsubsection{Merge and Simplify Rules}
\label{sec:algorithm:simplify}

\shortonly{This step attempts to merge and simplify rules using the same techniques as \cite{bui19mining}, extended with one additional simplification technique: If a subject/resource path of an atomic constraint evaluates to a same constant value $c$ for all of the subjects/resources that are in SRA tuples covered by the rule, then replace the atomic constraint with corresponding resource/subject condition and constant $c$.}

\fullonly{This step attempts to merge and simplify rules using the same techniques as \cite{bui19mining}.

First, this step attempts to merge pairs of rules that have the same subject type, resource type, and constraint by taking the least upper bound of their subject conditions, the least upper bound of their resource conditions, and the union of their sets of actions.  The {\em least upper bound} of conditions $c_1$ and $c_2$, denoted $c_1 \sqcup c_2$, is 
\begin{displaymath}
 \begin{array}{@{}l@{}}
 \{\tuple{p, {\rm in}, \val} \;|\; 
  \begin{array}[t]{@{}l@{}}
    (\exists \val_1,\val_2:\; \tuple{p, {\rm in}, \val_1} \in c_1 \land  \tuple{p, {\rm in}, \val_2} \in c_2\\
\ind\ind {} \land \val = \val_1 \union \val_2)\}
  \end{array}\\
{} \union \{\tuple{p, {\rm contains}, \val} \;|\; 
\begin{array}[t]{@{}l@{}}
  \tuple{p, {\rm contains}, \val} \in c_1\\
  {} \land\ \tuple{p, {\rm contains}, \val} \in c_2)\}.
\end{array}
 \end{array}
\end{displaymath}
When computing least upper bounds, DTRM$^-$  uses only positive atomic conditions; negative atomic conditions are dropped. Note that the meaning of the merged rule $\rhomerge$ is a superset of the meanings of the rules $\rho_1$ and $\rho_2$ being merged.  If the merged rule $\rhomerge$ is valid, then it replaces $\rho_1$ and $\rho_2$.

Second, this step attempts to simplify the rules as follows.
\begin{enumerate}

\item It eliminates atomic conditions from the subject and resource conditions when this preserves validity. Removing one atomic condition might prevent removal of another atomic condition, so it searches for a set of removable atomic conditions that maximizes the quality of the resulting rule.


\item It eliminates atomic constraints when this preserves validity.  It searches for the set of atomic constraints to remove that maximizes the quality of the resulting rule.

\item It eliminates overlapping actions between rules.  Specifically, an action $a$ in a rule $\rho$ is removed if there is another rule $\rho'$ in the policy such that $\scond(\rho') \subseteq \scond(\rho) \land \rcond(\rho') \subseteq \rcond(\rho) \land \con(\rho') \subseteq \con(\rho) \land a \in \acts(\rho')$.

\item It eliminates actions when this preserves the meaning of the policy. In other words, it removes an action $a$ in rule $\rho$ if all the SRA tuples covered by $a$ in $\rho$ are covered by other rules in the policy.  Note that the previous item is a special case of this one, listed separately to ensure that the special case takes precedence.

\item If the subject condition contains an atomic condition of the form $p=c$, and the constraint contains an atomic constraint of the form $p=p'$, then replace that atomic constraint with the atomic condition $p'=c$ in the resource condition (note that this is a form of constant propagation); and similarly for the symmetric situation in which the resource condition contains such an atomic condition, etc. 

DTRM$^-$ consider an additional case with the presence of negative condition/constraint. If the subject condition contains an atomic condition of the form $p=c$, and the constraint contains an atomic constraint of the form $p \neq p'$, then replace that atomic constraint with the atomic condition $p' \neq c$ in the resource condition; and similarly for the symmetric situation as mentioned in the first case.

%

\item Remove cycles in the paths in the conditions and constraint, if the resulting rule is valid and the resulting policy still covers all of $\spa$.  A cycle is a path that navigates from some class $C$ back to class $C$. 

\item If a subject/resource path of an atomic constraint evaluates to a same constant value $c$ for all of the subjects/resources that are in SRA tuples covered by the rule, then replace the atomic constraint with corresponding resource/subject condition and constant $c$. In DTRM$^-$, if the atomic constraint is negative, it will be replaced with the corresponding negative atomic condition. 
\end{enumerate}}

\arxivonly{
\subsection{Asymptotic Running Time}
\label{sec:asymptotic-running-time}


This section analyzes the asymptotic running time of our algorithm. We first analyze the complexity of phase 1. Let $n_{\rm feat}$ and $n_{\rm samp}$ be the number of features and feature vectors (samples), respectively; they depend on the size of the object model.
The cost of splitting samples at each node is $O(n_{\rm samp} \cdot n_{\rm feat})$; this is mainly the cost of computing the scoring criteria for each feature on the current samples set.  Let $sz_{\rm rule}$ be the ``size'' of the rules extracted from the tree, specifically, the sum of the numbers of features (conditions and constraints) in each extracted rule; typically, the size of these intermediate rules is comparable to the size of the final mined rules.  Note that the number of nodes in the tree is at most $sz_{\rm rule}$.  The cost of building the tree is $O(n_{\rm samp} \cdot n_{\rm feat} \cdot sz_{\rm rule})$, and the cost of extracting the rules is $O(sz_{\rm rule})$.  This cost for each tree is summed over the number of trees, which is the number of $\tuple{C_s, C_r, a}$ tuples, explained in Section \ref{sec:algorithm:dt}.


Now consider phase 2.  The running time of the eliminating negative features step depends on the running times of its substeps.  Recall that the substeps are applied in order until one succeeds.  The cost of checking whether a rule is valid and the cost of computing a rule's coverage are $O(n_{\rm samp})$, since they require assessing all samples.  Substep (1) takes $O(n_{\rm samp})$ time for the rule validity check, and $O(1)$ time for the negative feature removal.  Substep (2) takes $O(n_{\rm feat} \cdot n_{\rm samp})$ time to find a valid replacement feature among the set of all possible features. Let $n_{\rm obj}$ denote the maximum (over all types) number of objects of a single type in the object model. The cost of substep (3) is $O(n_{\rm obj})$; this is mainly the cost of computing the complement of the set of constants that appear in the negative atomic condition for a specify object type; recall that constants in our framework are object identifiers or boolean values.  Substep (4) takes $O(n_{\rm samp})$ time to compute the rule's coverage and extract the appropriate constants for the new atomic condition.  For substep (5), the worst-case cost is high, since it could need to consider all subsets of features, but in practice, this step is never even reached in our experiments (one of the first four steps always succeeds), so we omit it from the overall complexity analysis of the algorithm.  Let $n_{\rm neg}$ be the number of negative features generated from the first phase, which is typically small.  If the first 4 substeps are all applied for every negative feature,  the cost is $O(n_{\rm neg} \cdot ((n_{\rm feat} \cdot n_{\rm samp}) + n_{\rm obj}))$.


The running time of the merge rules step and simplify rules step depend on the number of rules generated in Phase 1. Let $n_{\rm rules}$ denotes the number of these rules; this is typically similar to the number of rules in the final mined policy. 
Let $n_{\rm cond}$ and $n_{\rm cons}$ be the maximum number of atomic conditions and atomic constraints, respectively, in each of these rules.  Let $lm$ (mnemonic for ``largest meaning'') denote the maximum value of $|\mean{\rho}|$ among all rules considered in these steps.  The value of $lm$ is at most $|\spa|$ but typically much smaller. 
The cost of checking rule validity in these stesp is $O(lm)$.

The cost of the merging step is $O(n_{\rm rules}^3 \cdot lm)$; note that the algorithm checks validity of merged rule for each merging attempt. The running time of the simplification step depends on its substeps.  The cost of substeps (1) and (2) are $O(n_{\rm rules} \cdot 2^{n_{\rm cond}} \cdot lm)$ and $O(n_{\rm rules} \cdot 2 ^ {n_{\rm cons}} \cdot lm)$, respectively; the exponential factors here are small in practice, because rules typically have only a few conditions and constraints (e.g., according to Table 1 in \cite{bui19mining}, the average number of conditions and constraints per rule in the policies considered there are at most 2.3 and 1.3, respectively).  The cost of each of substeps (3) and (4) is $O(n_{\rm rules}^2 \cdot |\Act| \cdot lm)$.  The cost of substep (5) is $O(n_{\rm rules} \cdot n_{\rm cond} \cdot n_{\rm cons})$.  The cost of substep (6) is $O(n_{\rm rules} \cdot (n_{\rm cond} + n_{\rm cons}) \cdot n_{\rm class})$, where $n_{\rm class}$ is the number of classes in the class model. The cost of substep (7) is $O(n_{\rm rules} \cdot n_{\rm cons} \cdot lm)$.
}

\section{Evaluation Methodology}
\label{sec:evaluation-methodology}

\begin{figure*}[htb]
\begin{tabular}[b]{|l|l|l|l|l|}
\hline
\multicolumn{1}{|c|}{Policy\_N} & \multicolumn{1}{c|}{\#obj} & \multicolumn{1}{c|}{\#field} & \multicolumn{1}{c|}{\#FV} & \multicolumn{1}{c|}{\#rule} \\ \hline
EMR\_15 & 353 & 877 & 4134 & 6 \\ \hline
healthcare\_5 & 736 & 1804 & 42121 & 8 \\ \hline
project-mgmt\_5 & 179 & 296 & 4080 & 10 \\ \hline
university\_5 & 738 & 926 & 83761 & 10 \\ \hline
e-document\_75 & 284 & 1269 & 31378 & 39 \\ \hline
e-document\_100 & 352 & 1653 & 52466 & 39 \\ \hline
e-document\_125 & 423 & 2065 & 82860 & 39 \\ \hline
e-document\_150 & 486 & 2406 & 108403 & 39 \\ \hline
e-document\_175 & 563 & 2830 & 152093 & 39 \\ \hline
eWorkforce\_10 & 412 & 1124 & 14040 & 19 \\ \hline
eWorkforce\_15 & 585 & 1647 & 31769 & 19 \\ \hline
eWorkforce\_20 & 691 & 1963 & 45625 & 19 \\ \hline
eWorkforce\_25 & 862 & 2484 & 74856 & 19 \\ \hline
eWorkforce\_30 & 1016 & 2928 & 104845 & 19 \\ \hline
syn\_20\_$x$ & 678 & 7848 & 25600 & $x$ \\ \hline
syn\_25\_20 & 828 & 9773 & 40000 & 20 \\ \hline
syn\_30\_20 & 978 & 11698 & 57600 & 20 \\ \hline
syn\_35\_20 & 1128 & 13623 & 78400 & 20 \\ \hline
\end{tabular}
\hspace*{0.5em}\acmonly{\includegraphics[width=0.5\textwidth]{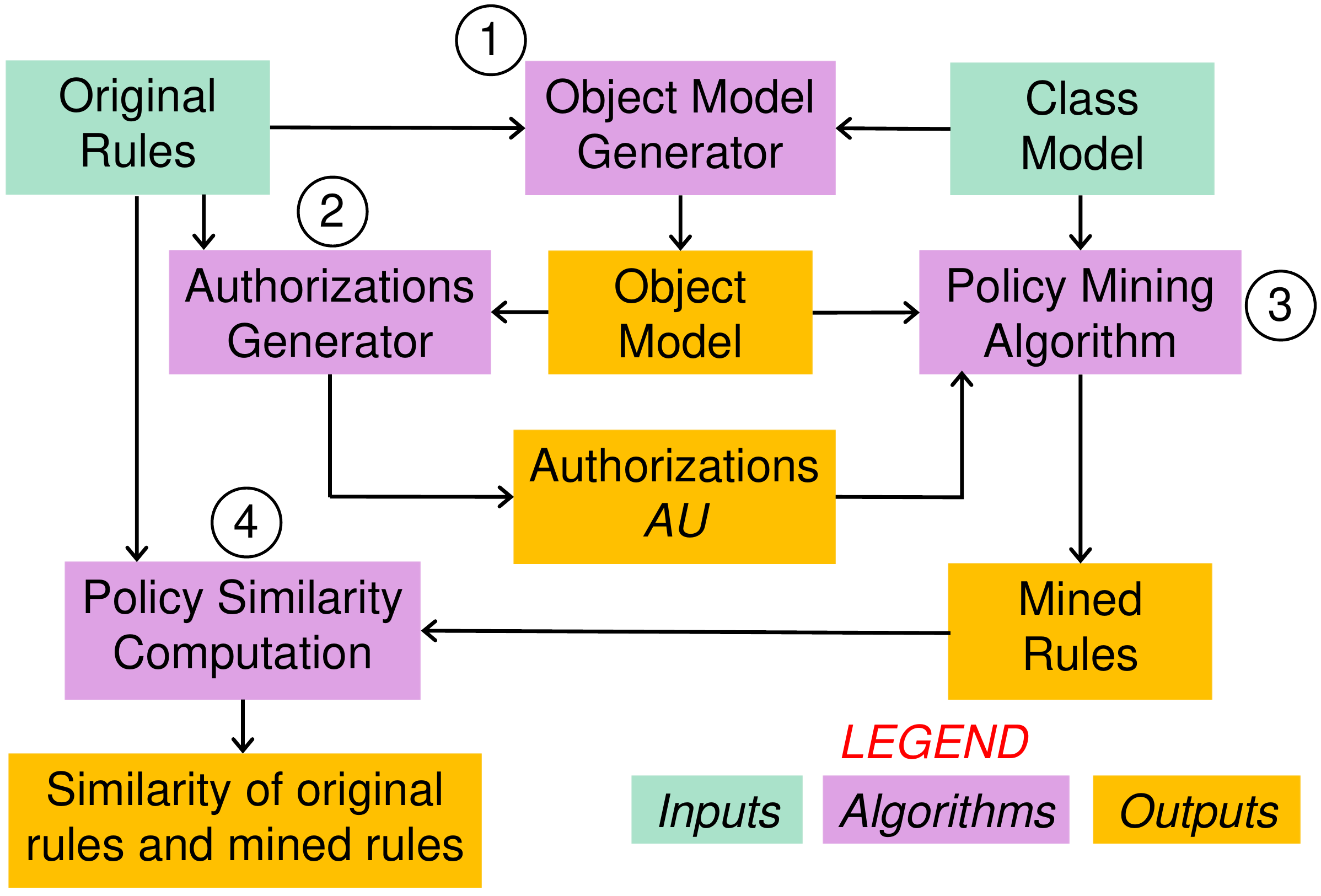}}\arxivonly{\includegraphics[width=0.45\textwidth]{eval-methodology.pdf}}
\smallskip
\caption{Left: Policy sizes. For the given value of the object model size parameter $N$, \#obj is the average number of objects in the object model, and \#field is the average number of fields in the object model, i.e., the sum over objects $o$ of the number of fields in $o$.  \#FV is the number of feature vectors (i.e., labeled SRA tuples) that the algorithms use to train a classifier.  Averages are over 5 pseudorandom object models for each policy. For the syn\_20\_$M$ policies, the number of rules $M$ is 10, 20, 30 or 40. Right: Evaluation methodology; reproduced from \cite{bui19sacmat}.}
\label{tab:policy-size}
\label{fig:eval-method}
\end{figure*}


We adopt Bui et al.'s methodology for evaluating policy mining algorithms \cite{bui19sacmat}.  It is depicted in Figure \ref{fig:eval-method}.  It takes a class model and a set of ReBAC rules as inputs.  The methodology is to generate an object model based on the class model (independent of the ReBAC rules), compute the authorizations $\spa$ from the object model and the rules, run the policy mining algorithm with the class model, object model, and $\spa$ as inputs, and finally compare the mined policy rules with the simplified original (input) policy rules, obtained by applying the simplifications in Section \ref{sec:algorithm:simplify} to the given rules.  Comparison with the simplified original policy is a more robust measure of the algorithm's ability to discover high-level rules than comparison with the original policy, because the original policy is not always the simplest.  If the mined rules are similar to the simplified original rules, the policy mining algorithm succeeded in discovering the desired ReBAC rules that are implicit in $\spa$.


\subsection{Datasets}
\label{sec:datasets}

We use four sample policies developed by Bui et al. \cite{bui19mining}.  One is for electronic medical records (EMR), based on the EBAC policy in \cite{bogaerts15entityShort}, translated to ReBAC; the other three are for healthcare, project management, and university records, based on ABAC policies in \cite{xu15miningABAC}, generalized and made more realistic, taking advantage of ReBAC's expressiveness.  These  policies are non-trivial but relatively small.

We also use Bui et al.'s translation into ORAL2 \cite{bui19mining} of two large case studies developed by Decat, Bogaerts, Lagaisse, and Joosen based on the access control requirements for Software-as-a-Service (SaaS) applications offered by real companies \cite{decat14edoc,decat14workforce}.  One is for a SaaS multi-tenant e-document processing application; the other is for a SaaS workforce management application provided by a company that handles the workflow planning and supply management for product or service appointments (e.g., install or repair jobs).

Finally, we use the synthetic ORAL2 policies described in \cite{bui19sacmat} and some extensions of them with additional rules.

All of the object models are generated by policy-specific pseudorandom algorithms designed to produce realistic object models, by creating objects and selecting their attribute values using appropriate probability distributions.  These algorithms are parameterized by a size parameter $N$\fullonly{; for most classes, the number of instances is selected from a normal distribution whose mean is linear in $N$}.  \fullonly{Bui et al.'s policy rules and object model generators for the sample policies and case studies, and their synthetic policy generator, are available online \cite{stollersoftware}.}\shortonly{Bui et al.'s policy rules, object model generators, and synthetic policy generator are available online \cite{stollersoftware}.}  We slightly modified the object model generators for the project management, workforce management, and e-document policies, to make the generated object models slightly more realistic.  More details about object model generation are in \cite{bui19sacmat,bui19mining}.

These policies, or variants of them, have been used as benchmarks in several other papers on policy mining\shortonly{, including \cite{iyer2019,bui18mining,medvet2015,iyer2018}}.\fullonly{  In work on ReBAC mining, Iyer et al. \cite{iyer2019} use variants of parts of three of the sample policies and the workforce management case study, and Bui et al. \cite{bui18mining} use all of the sample policies and case studies.  In work on ABAC mining, Medvet et al. \cite{medvet2015}, Iyer at al. \cite{iyer2018}, and Karimi et al. \cite{karimi2018} use Xu et al.'s original ABAC versions of some of the sample policies.}

The table in Figure \ref{tab:policy-size} shows several metrics of the size of the rules, class model, and object model in each policy. \#field is computed by summing, over the objects in the object model, the number of fields (including ``id'' field and Boolean fields) in each object.



\fullonly{
The {\em Electronic Medical Record (EMR) sample policy}, based on the EBAC policy in \cite{bogaerts15entity}, controls access by physicians and patients to electronic medical records, based on institutional affiliations, patient-physician consultations (each EMR is associated with a consultation), supervisor relationships among physicians, etc. The numbers of physicians, consultations, EMRs, and hospitals are proportional to $N$.

The {\em healthcare sample policy}, based on the ABAC policy in \cite{xu15miningABAC}, controls access by nurses, doctors, patients, and agents (e.g., a patient's spouse) to electronic health records (HRs) and HR items (i.e., entries in health records). The numbers of wards, teams, doctors, nurses, teams, patients, and agents are proportional to $N$.

The {\em project management sample policy}, based on the ABAC policy in \cite{xu15miningABAC}, controls access by department managers, project leaders, employees, contractors, auditors, accountants, and planners to budgets, schedules, and tasks associated with projects. The numbers of departments, projects, tasks, and users of each type are proportional to $N$.  

The {\em university sample policy}, based on the ABAC policy in \cite{xu15miningABAC}, controls access by students, instructors, teaching assistants (TAs), department chairs, and staff in the registrar's office and admissions office to applications (for admission), gradebooks, transcripts, and course schedules. The numbers of departments, students, faculty, and applicants for admission are proportional to $N$.


The {\em e-document case study}, based on \cite{decat14edoc}, is for a SaaS multi-tenant e-document processing application.  The application allows tenants to distribute documents to their customers, either digitally or physically (by printing and mailing them). The overall policy contains rules governing document access and administrative operations by employees of the e-document company, such as helpdesk operators and application administrators.  It also contains specific policies for some sample tenants. One sample tenant is a large bank, which controls permissions to send and read documents based on (1) employee attributes such as department and projects, (2) document attributes such as document type, related project (if any), and presence of confidential or personal information, and (3) the bank customer to which the document is being sent.Some tenants have semi-autonomous sub-organizations, modeled as sub-tenants, each with its own specialized policy rules. The numbers of employees of each tenant, registered users of each customer organization, and documents are proportional to $N$.  

The {\em workforce management case study}, based on \cite{decat14workforce}, is for a SaaS workforce management application provided by a company, pseudonymously called eWorkforce, that handles the workflow planning and supply management for product or service appointments (e.g., install or repair jobs).  Tenants (i.e., eWorkforce customers) can create tasks on behalf of their customers.
Technicians working for eWorkforce, its workforce suppliers, or subcontractors of its workforce suppliers receive work orders to work on those tasks, and appointments are scheduled if appropriate.  Warehouse operators receive requests for required supplies. The overall policy contains rules governing the employees of eWorkforce, as well as specific policies for some sample tenants, including PowerProtection (a provider of power protection equipment and installation and maintenance services) and TelCo (a telecommunications provider, including installation and repair services).  Permissions to view, assign, and complete tasks are based on each subject's position, the assignment of tasks to technicians, the set of technicians each manager supervises, the contract (between eWorkforce and a tenant) that each work order is associated with, the assignment of contracts to departments within eWorkforce, etc. The only change we make is to omit from the workforce management case study the classes and 7 rules related to work orders, because they involve inheritance, which our algorithm does not yet support (it is future work).  The numbers of helpdesk suppliers, workforce providers, subcontractors, helpdesk operators, contracts, work orders, etc., are proportional to $N$.


The {\em synthetic policies} developed by Bui et al. \cite{bui19sacmat} are designed to have realistic structure, statistically similar in some ways to the sample policies and case studies described above.  The class model is designed to allow generating atomic conditions and atomic constraints with many combinations of path length and operator.  It supports the types of conditions and constraints that appear in the sample policies and case studies, plus constraints involving the additional constraint operators that are supported in ORAL2 but not in the original ORAL \cite{bui19mining}.  The object model generator's size parameter $N$ specifies the desired number of instances of each subject class.  The number of instances of each resource class is $5 \cdot N$.  The numbers of instances of other classes is fixed at 3.  This reflects a typical structure of realistic policies, in which the numbers of instances of some classes (e.g., doctors, patients, health records) scale linearly with the overall size of the organization, while the numbers of instances of other classes (e.g., departments, medical specialties) grow much more slowly (which we approximate as constant).\arxivonly{  Policy rules are generated using several numbers and statistical distributions based on the rules in the sample policies and case studies.}
}

\subsection{Policy Similarity Metrics}
\label{sec:policy-sim-metrics}

We evaluate the quality of the generated policy primarily by its {\em syntactic similarity} and {\em policy semantic similarity} to the simplified original policy.  These metrics are first defined in \cite{xu15miningABAC,bui18mining} and are normalized to range from 0 (completely different) to 1 (identical).  We adapt the syntactic similarity metric to take negation into account.  The metrics are based on Jaccard similarity of sets, defined by $J(S_1, S_2) = |S_1\intersect S_2| \,/\, |S_1 \union S_2|$.  For convenience, we extend $J$ to apply to single values: $J(v_1,v_2)$ is 1 if $v_1=v_2$ and 0 otherwise.

{\em Syntactic similarity} of policies measures the syntactic similarity of rules in the policies, based on the fractions of types, conditions, constraints, and actions that rules have in common. The {\em syntactic similarity of rules} is defined bottom-up as follows.    
For a possibly negated atomic condition $ac$, let $\getsign(ac)$, $\getpath(ac)$, and $\getval(ac)$ denote its sign (positive or negative), its path, and its value (or set of values), respectively.  Syntactic similarity of atomic conditions $ac_1$ and $ac_2$ is 0 if they contain different paths, otherwise it is the mean of
 $J(\getsign(ac_1), \getsign(ac_2))$, $J(\getpath(ac_1), \getpath(ac_2))$,
and $J(\getval(ac_1), \getval(ac_2)))$; we do not explicitly compare the operators, because atomic conditions with the same path must have the same operator, since the operator is uniquely determined by the multiplicity of the path.  For a set $S$ of atomic conditions, let $\paths(S)= \{ \getpath(ac)\;|\; ac \in S \}$.  For sets $S_1$ and $S_2$ of atomic conditions,
\begin{displaymath}
\synSim(S_1, S_2) = |\paths(S_1) \union \paths(S_2)|^{-1} \acmonly{\!}\sum_{ac_1 \in S_1, ac_2 \in S_2} \acmonly{\!\!}\acSim(ac_1, ac_2)
\end{displaymath}
The syntactic similarity of rules $\rho_1=\langle st_1, sc_1, rt_1,$ $rc_1, c_1, A_1\rangle$ and $\rho_2=\tuple{st_2, sc_2, rt_2, rc_2, c_2, A_2}$ is
$\synSim(\rho_1, \rho_2) = \average(%
    J(st_1, st_2), 
    \synSim(sc_1,$ $sc_2),
    J(rt_1, rt_2),
    \synSim(rc_1, rc_2), J(c_1, c_2),  J(A_1, A_2))$.


The {\em syntactic similarity of policies} $\pi_1$ and $\pi_2$, $\synSim(\pi_1$, $\pi_2$), is the average, over rules $\rho$ in $\pi_1$, of the syntactic similarity between $\rho$ and the most similar rule in $\pi_2$. 

The {\em semantic similarity of polices} measures the fraction of authorizations that the policies have in common.  Specifically, the {\em semantic similarity} of policies $\pi_1$ and $\pi_2$ is $J(\mean{\pi_1},\mean{\pi_2})$. 

\section{Evaluation Results}
\label{sec:evaluation-results}

\begin{table*}[htb]
\resizebox{\textwidth}{!}{%
\begin{tabular}{|l|l|l|l|l|l|l|l|l|l|l|l|l|l|l|}
\hline
\multicolumn{1}{|c|}{\multirow{3}{*}{Policy}} & \multicolumn{6}{c|}{Syntactic Similarity} & \multicolumn{8}{c|}{Running Time (sec)} \\ \cline{2-15} 
\multicolumn{1}{|c|}{} & \multicolumn{2}{c|}{FS-SEA*} & \multicolumn{2}{c|}{DTRM} & \multicolumn{2}{c|}{DTRM$^-$} & \multicolumn{2}{c|}{FS-SEA*} & \multicolumn{2}{c|}{DTRM} & \multicolumn{1}{c|}{\multirow{2}{*}{SpdUp}} & \multicolumn{2}{c|}{DTRM$^-$} & \multicolumn{1}{c|}{\multirow{2}{*}{SpdUp}} \\ \cline{2-11} \cline{13-14}
\multicolumn{1}{|c|}{} & $\mu$ & $\sigma$ & $\mu$ & $\sigma$ & $\mu$ & $\sigma$ & $\mu$ & $\sigma$ & $\mu$ & $\sigma$ & \multicolumn{1}{c|}{} & $\mu$ & $\sigma$ & \multicolumn{1}{c|}{} \\ \hline
EMR\_15 & 0.99 & 0.01 & 0.99 & 0.01 & 0.99 & 0.01 & 96 & 7.37 & 56 & 0.30 & 1.70 & 53 & 5.55 & 1.82 \\ \hline
healthcare\_5 & 1.00 & 0.00 & 1.00 & 0.00 & 1.00 & 0.00 & 111 & 14.54 & 80 & 41.07 & 1.39 & 70 & 29.68 & 1.57 \\ \hline
project-mgmt.\_5 & 1.00 & 0.00 & 1.00 & 0.00 & 1.00 & 0.00 & 6 & 0.45 & 2 & 1.62 & 3.88 & 2 & 0.39 & 4.07 \\ \hline
university\_5 & 1.00 & 0.00 & 1.00 & 0.00 & 1.00 & 0.00 & 271 & 21.98 & 159 & 64.02 & 1.70 & 131 & 44.32 & 2.07 \\ \hline
e-doc.\_75 & 0.93 & 0.02 & 0.90 & 0.01 & 0.90 & 0.01 & 696 & 133.88 & 296 & 42.57 & 2.35 & 121 & 14.90 & 5.75 \\ \hline
e-doc.\_100 & 0.94 & 0.01 & 0.91 & 0.03 & 0.90 & 0.02 & 1734 & 542.88 & 650 & 64.94 & 2.67 & 250 & 19.14 & 6.94 \\ \hline
e-doc.\_125 & 0.93 & 0.01 & 0.92 & 0.01 & 0.93 & 0.02 & 3516 & 1415.93 & 1200 & 276.13 & 2.93 & 481 & 24.52 & 7.31 \\ \hline
e-doc.\_150 & 0.91 & 0.01 & 0.94 & 0.01 & 0.94 & 0.00 & 6068 & 1202.25 & 2292 & 323.12 & 2.65 & 735 & 58.44 & 8.25 \\ \hline
e-doc.\_175 & 0.92 & 0.01 & 0.93 & 0.01 & 0.93 & 0.01 & 15218 & 4535.45 & 3823 & 460.36 & 3.98 & 1227 & 68.07 & 12.41 \\ \hline
eWorkforce\_10 & 0.97 & 0.01 & 0.98 & 0.01 & 0.97 & 0.01 & 70 & 9.32 & 50 & 5.80 & 1.41 & 48 & 0.69 & 1.47 \\ \hline
eWorkforce\_15 & 0.95 & 0.02 & 0.98 & 0.02 & 0.97 & 0.02 & 287 & 37.68 & 182 & 13.62 & 1.58 & 176 & 2.06 & 1.63 \\ \hline
eWorkforce\_20 & 0.92 & 0.03 & 0.98 & 0.02 & 0.96 & 0.02 & 669 & 89.99 & 426 & 94.17 & 1.57 & 319 & 5.16 & 2.10 \\ \hline
eWorkforce\_25 & 0.95 & 0.04 & 0.97 & 0.02 & 0.95 & 0.02 & 1750 & 294.25 & 946 & 280.87 & 1.85 & 745 & 10.67 & 2.35 \\ \hline
eWorkforce\_30 & 0.97 & 0.02 & 0.97 & 0.02 & 0.95 & 0.02 & 3113 & 725.28 & 1492 & 312.45 & 2.09 & 1378 & 17.42 & 2.26 \\ \hline
syn\_20\_10 & 0.99 & 0.00 & 0.99 & 0.01 & 0.99 & 0.01 & 938 & 348.24 & 166 & 52.16 & 5.66 & 142 & 12.42 & 6.61 \\ \hline
syn\_20\_20 & 0.98 & 0.01 & 0.98 & 0.02 & 0.97 & 0.04 & 3129 & 887.18 & 309 & 88.93 & 10.11 & 256 & 24.97 & 12.22 \\ \hline
syn\_20\_30 & 0.99 & 0.00 & 0.99 & 0.01 & 0.98 & 0.03 & 6303 & 1258.03 & 379 & 88.73 & 16.65 & 317 & 27.25 & 19.86 \\ \hline
syn\_20\_40 & 0.99 & 0.01 & 0.98 & 0.02 & 0.97 & 0.03 & 11169 & 2812.94 & 435 & 69.75 & 25.67 & 370 & 14.52 & 30.21 \\ \hline
syn\_25\_20 & 1.00 & 0.00 & 0.98 & 0.02 & 0.97 & 0.04 & 6494 & 2283.31 & 571 & 142.31 & 11.38 & 485 & 42.43 & 13.40 \\ \hline
syn\_30\_20 & 1.00 & 0.00 & 0.99 & 0.01 & 0.99 & 0.01 & 11161 & 3396.60 & 898 & 82.72 & 12.43 & 861 & 75.63 & 12.96 \\ \hline
syn\_35\_20 & 0.99 & 0.01 & 0.99 & 0.01 & 0.99 & 0.01 & 21758 & 7355.68 & 1419 & 138.17 & 15.33 & 1416 & 114.47 & 15.36 \\ \hline
\end{tabular}%
}\smallskip
\caption{Comparison of DTRM, DTRM$^-$, and FS-SEA*. 
$\mu$ and $\sigma$ are the mean and standard deviation, respectively. SpdUp is the speed up, computed as the ratio of the running time of each of our algorithms to the running time of FS-SEA*. 
}
\label{tab:syn-sim-running-time}
\end{table*}

\begin{figure*}[tbp]
\begin{tabular}[t]{@{}l@{}}
  \centering
\includegraphics[width=1.0\textwidth]{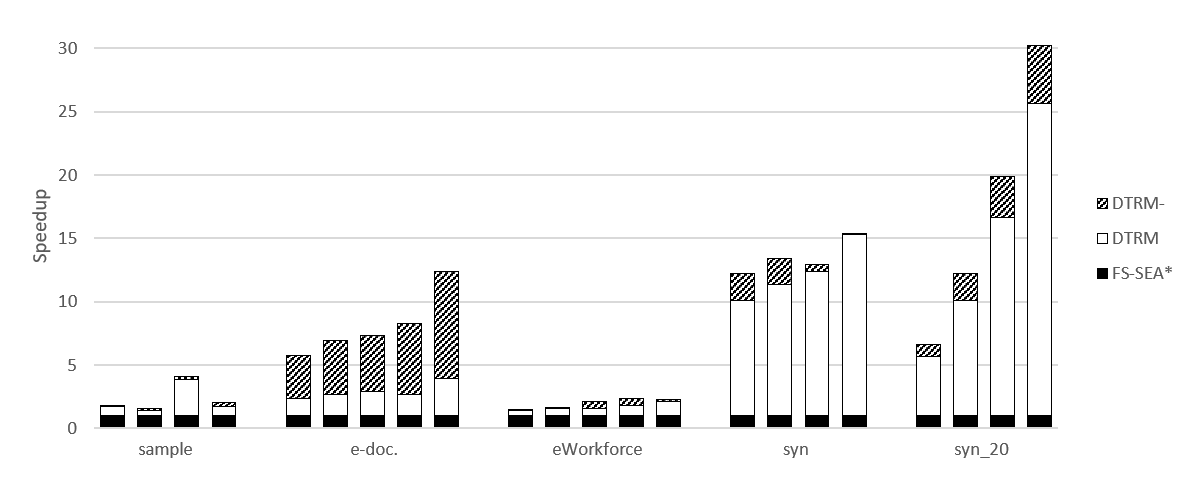}
\end{tabular}
  \caption{Speedups of DTRM and DTRM$^-$ relative to FS-SEA*.  There are 5 clusters, corresponding to 5 groups of policies. The ``sample'' cluster contains bars for the following policies (from left to right): EMR\_15, healthcare\_5, project-management\_5 and university\_5;  ``e-doc'' cluster for e-document\_75, e-document\_100, e-document\_125, e-document\_150, e-document\_175; ``eWorkforce'' cluster for eWorkforce\_10, eWorkforce\_15, eWorkforce\_20, eWorkforce\_25, eWorkforce\_30; ``syn'' cluster for syn\_20\_20, syn\_25\_20, syn\_30\_20, syn\_35\_20; ``syn\_20'' cluster for syn\_20\_10, syn\_20\_20, syn\_20\_30, syn\_20\_40.}
  \label{fig:running-time}
\end{figure*}
This section presents the results of experiments comparing our algorithms with Bui et al.'s  FS-SEA* algorithm \cite{bui19sacmat} and Iyer et al.'s algorithm \cite{iyer2019}.  DTRM and DTRM$^-$ are implemented in Python, except that phase 2 step 2 (merge and simplify rules) uses the Java code from Bui et al.'s implementation of FS-SEA*, available at \cite{stollersoftware}.  Experiments were run on Windows 10 on an Intel i7-6770HQ CPU. 
In summary, we find that: (1) compared with FS-SEA*, our algorithms are comparably effective at discovering the desired ReBAC rules, and are significantly faster, with the speedup exceeding $10\times$ for several datasets and generally increasing with policy size, hence expected be even larger for the large datasets arising in practice; and (2) compared with Iyer et al.'s algorithm, our algorithms are several times faster, and produce policies that are the same size or smaller (fewer rules) and more similar to the original policies.

\subsection{Comparison with FS-SEA*}
\label{sec:evaluation-results:compare-FS-SEA}


We compared DTRM and DTRM$^-$ with FS-SEA* using the datasets described in Section \ref{sec:datasets}.  We use the same path length limits ({\it cf.} Section \ref{sec:algorithm:dt}) as in \cite{bui19mining,bui19sacmat}.  For the case studies, we generated policies with varying size (of the object model): $N = 10, 15, 20, 25, 30, 35$ for eWorkforce and $N = 75, 100, 125, 150, 175$ for e-document. For each size, we generated 5 pseudo-random object models. For synthetic policies, we generated two families of policies.  Synthetic policies are designated by syn\_$N$\_$M$, where $N$ is the object model size parameter, and $M$ is the number of rules.  The first family consists of 5 sets of $M=20$ synthetic rules, and object models with sizes $N = 20, 25, 30$ (one of each size); we chose $M=20$ because it is the average number of rules in the sample policies and case studies.  The second family consists of sets of $M=10, 30, 40$ synthetic rules (one of each size), and 5 object models with size $N=20$. We ran DTRM, DTRM$^-$, and FS-SEA* on all of them, and average the results for the five policies with the same $N$ and $M$.  The standard deviations are reasonable, indicating that averaging over 5 object models for each data point is sufficient to obtain meaningful results.





\subsubsection{Policy Similarity and WSC}

All three algorithms always mine policies that grant exactly the same authorizations as the input ACL policies and thus achieve perfect {\em semantic similarity} for all datasets.

All algorithms achieve similar {\em syntactic similarity} when comparing mined rules with simplified original rules, as explained in Section \ref{sec:evaluation-methodology}.  The minimum, median, and maximum (over all datasets) syntactic similarity achieved by each algorithm are: 0.91, 0.98, 1.0 for FS-SEA*; 0.90, 0.98, 1.0 for DRTM; and 0.90, 0.97, 1.0 for DTRM$^-$. The syntactic similarity achieved by DTRM and DTRM- are usually the same or better than that achieved by FS-SEA*, and in the worst cases in Table \ref{tab:syn-sim-running-time}, are at most 2\% and 4\% lower, respectively.
DTRM$^-$ achieved slightly lower syntactic similarity since the input policies do not use any negative atomic condition/constraint.

We report results for WSC in terms of the ratio of the WSC of the policy mined by DTRM or DTRM$^-$ to the WSC of the policy mined by FS-SEA*; thus a ratio below 1 means that DTRM or DTRM$^-$ produce a more concise policy than FS-SEA*. The minimum, median, and maximum (over all datasets) of this ratio are: 0.79, 1.0, 1.21 for DRTM, and 0.86, 1.01, 1.32 for DTRM$^-$. WSC of policies mined by DTRM$^-$ is not smaller than WSC of policies mined by DTRM, even though theoretically negation could allow more concise policies.  This indicates that DTRM$^-$ sometimes produces policies that use negation even when it is not beneficial.  This is not surprising, because when constructing the decision tree, the algorithm does not have a preference for using or avoiding negation.

\arxivonly{
\begin{table*}[thb]
\resizebox{\textwidth}{!}{%
\begin{tabular}{|l|l|l|l|l|l|l|l|l|l|l|l|l|l|l|l|l|l|l|}
\hline
\multicolumn{1}{|c|}{\multirow{3}{*}{Policy}} & \multicolumn{11}{c|}{WSC} & \multicolumn{7}{c|}{Number of Rules} \\ \cline{2-19} 
\multicolumn{1}{|c|}{} & \multicolumn{1}{c|}{Orig} & \multicolumn{2}{c|}{sOrig} & \multicolumn{2}{c|}{FS-SEA*} & \multicolumn{2}{c|}{DTRM} & \multicolumn{1}{c|}{\multirow{2}{*}{Ratio}} & \multicolumn{2}{c|}{DTRM$^-$} & \multicolumn{1}{c|}{\multirow{2}{*}{Ratio}} & \multicolumn{1}{c|}{Orig} & \multicolumn{2}{c|}{FS-SEA*} & \multicolumn{2}{c|}{DTRM} & \multicolumn{2}{c|}{DTRM$^-$} \\ \cline{2-8} \cline{10-11} \cline{13-19} 
\multicolumn{1}{|c|}{} & $\mu$ & $\mu$ & $\sigma$ & $\mu$ & $\sigma$ & $\mu$ & $\sigma$ & \multicolumn{1}{c|}{} & $\mu$ & $\sigma$ & \multicolumn{1}{c|}{} & $\mu$ & $\mu$ & $\sigma$ & $\mu$ & $\sigma$ & $\mu$ & $\sigma$ \\ \hline
EMR\_15 & 49 & 47 & 0.00 & 47 & 0.00 & 47 & 0.00 & 1.00 & 47 & 0.00 & 1.00 & 6 & 6 & 0.00 & 6 & 0.00 & 6 & 0.00 \\ \hline
healthcare\_5 & 39 & 39 & 0.00 & 39 & 0.00 & 39 & 0.00 & 1.00 & 39 & 0.00 & 1.00 & 8 & 8 & 0.00 & 8 & 0.00 & 8 & 0.00 \\ \hline
project-mgmt.\_5 & 67 & 67 & 0.00 & 67 & 0.00 & 67 & 0.00 & 1.00 & 67 & 0.00 & 1.00 & 10 & 10 & 0.00 & 10 & 0.00 & 10 & 0.00 \\ \hline
university\_5 & 54 & 54 & 0.00 & 54 & 0.00 & 54 & 0.00 & 1.00 & 54 & 0.00 & 1.00 & 10 & 10 & 0.00 & 10 & 0.00 & 10 & 0.00 \\ \hline
e-doc.\_75 & 359 & 248 & 1.96 & 252 & 6.09 & 305 & 21.27 & 1.21 & 325 & 18.34 & 1.29 & 39 & 32.2 & 0.45 & 39 & 2.17 & 40 & 1.79 \\ \hline
e-doc.\_100 & 359 & 251 & 2.53 & 255 & 3.67 & 309 & 26.95 & 1.21 & 337 & 22.26 & 1.32 & 39 & 32 & 0.71 & 39 & 3.32 & 40 & 2.30 \\ \hline
e-doc.\_125 & 359 & 252 & 1.67 & 273 & 8.90 & 302 & 16.39 & 1.10 & 321 & 26.91 & 1.17 & 39 & 33.2 & 1.09 & 38 & 1.76 & 38 & 2.28 \\ \hline
e-doc.\_150 & 359 & 251 & 1.79 & 278 & 13.84 & 277 & 7.28 & 0.99 & 300 & 5.95 & 1.08 & 39 & 33.6 & 1.52 & 36 & 0.89 & 36 & 0.89 \\ \hline
e-doc.\_175 & 359 & 251 & 0.80 & 317 & 49.07 & 297 & 26.57 & 0.94 & 322 & 30.96 & 1.02 & 39 & 34.4 & 1.82 & 38 & 3.44 & 38 & 3.05 \\ \hline
eWorkforce\_10 & 171 & 133 & 7.30 & 157 & 8.79 & 128 & 9.09 & 0.81 & 136 & 10.27 & 0.87 & 19 & 18 & 0.00 & 19 & 0.45 & 20 & 0.45 \\ \hline
eWorkforce\_15 & 171 & 131 & 8.97 & 154 & 2.87 & 125 & 2.60 & 0.82 & 135 & 5.64 & 0.88 & 19 & 16.8 & 1.09 & 18 & 0.55 & 19 & 0.55 \\ \hline
eWorkforce\_20 & 171 & 131 & 5.82 & 159 & 5.60 & 131 & 5.42 & 0.83 & 139 & 6.66 & 0.88 & 19 & 17.2 & 1.09 & 19 & 0.55 & 20 & 0.55 \\ \hline
eWorkforce\_25 & 171 & 130 & 6.95 & 163 & 12.10 & 129 & 13.99 & 0.79 & 140 & 16.58 & 0.86 & 19 & 18.2 & 0.45 & 19 & 1.09 & 20 & 1.10 \\ \hline
eWorkforce\_30 & 171 & 129 & 1.20 & 165 & 5.45 & 143 & 13.76 & 0.87 & 151 & 15.60 & 0.92 & 19 & 18.2 & 1.09 & 19 & 1.52 & 20 & 1.52 \\ \hline
syn\_20\_10 & 73 & 71 & 3.66 & 71 & 3.65 & 70 & 3.12 & 0.99 & 72 & 6.42 & 1.02 & 10 & 9.8 & 0.45 & 10 & 0.71 & 10 & 0.71 \\ \hline
syn\_20\_20 & 155 & 150 & 7.29 & 169 & 29.65 & 156 & 9.20 & 0.92 & 161 & 9.82 & 0.95 & 20 & 20.2 & 0.84 & 21 & 0.89 & 21 & 1.41 \\ \hline
syn\_20\_30 & 230 & 211 & 6.76 & 217 & 5.46 & 214 & 9.56 & 0.99 & 220 & 14.00 & 1.02 & 30 & 28.8 & 1.09 & 29 & 1.41 & 29 & 1.95 \\ \hline
syn\_20\_40 & 312 & 280 & 11.48 & 282 & 6.91 & 286 & 18.25 & 1.01 & 292 & 24.02 & 1.03 & 40 & 37.4 & 0.55 & 38 & 2.07 & 39 & 2.77 \\ \hline
syn\_25\_20 & 155 & 150 & 7.29 & 150 & 7.29 & 156 & 6.60 & 1.04 & 161 & 7.98 & 1.07 & 20 & 19.60 & 0.55 & 20 & 1.14 & 21 & 1.58 \\ \hline
syn\_30\_20 & 155 & 150 & 7.29 & 150 & 7.29 & 153 & 5.89 & 1.02 & 153 & 6.58 & 1.02 & 20 & 19.60 & 0.55 & 20 & 1.30 & 20 & 1.30 \\ \hline
syn\_35\_20 & 155 & 150 & 7.29 & 150 & 7.17 & 151 & 6.65 & 1.01 & 152 & 7.57 & 1.01 & 20 & 19.60 & 0.55 & 20 & 1.22 & 20 & 1.22 \\ \hline
\end{tabular}%
}\smallskip
\caption{Comparison of DTRM, DTRM$^-$, and FS-SEA*:   WSC and number of rules in mined policies.  Orig and sOrig refer to the original rules and the simplified original rules, respectively.  Ratio is the ratio of the WSC of the policy mined by DTRM or DTRM$^-$ to the WSC of the policy mined by FS-SEA*. 
}
\label{tab:wsc-num-rules}
\end{table*}}

Detailed results for policy similarity appear in Table \ref{tab:syn-sim-running-time}.\acmonly{  Detailed results for WSC appear in \cite{DTRM-arxiv}.}\arxivonly{  Detailed results for WSC appear in Table \ref{tab:wsc-num-rules}.}
%
We conclude that all three algorithms produce policies with similar quality according to all three metrics.

\subsubsection{Running Time}

We report results for running time as the speedup relative to FS-SEA*, i.e., the ratio of the running time of each algorithm to the running time of FS-SEA*.\fullonly{  Detailed results appear in Table \ref{tab:syn-sim-running-time}.}  The results are summarized in the stacked bar chart in Figure \ref{fig:running-time}.  Each bar has three segments, representing three overlaid bars, each corresponding to an algorithm.  The total height (as measured on the y-axis) of the top of each segment is the speedup of that algorithm.  The first (black) segment is for FS-SEA*, so it always has height 1. The second (white) segment is for DTRM. The third (shaded) segment is for DTRM$^-$. For example, if DTRM achieved speedup 2.2 and DTRM$^-$ achieved speedup 4.4 for some policy, then the top of the black segment would be at height 1, the top of the white segment at height 2.2 (hence the white segment would be 1.2 units long), and the top of the shaded segment at height 4.4.  This stacked bar chart format is suitable for reporting the speedups because, in all of our experiments, DTRM$^-$ is faster than DTRM, and DTRM is faster than FS-SEA*.  The bars within each cluster other than the sample policy cluster are ordered left-to-right by increasing policy size, specifically by object model size for the e-doc., eWorkforce, and syn clusters, and by number of rules for the syn\_20 cluster.  Observe that speedup generally increases from left to right within those clusters, i.e., generally increases with policy size. A main reason that speedups for e-document and synthetic policies are larger than for eWorkforce and most sample policies is that the former policies have a larger number of rules per $\tuple{C_s, C_r, a}$ tuple (explained in Section \ref{sec:algorithm:dt}). DTRM (and DTRM$^-$) achieve larger speedups for such policies, because FS-SEA* repeats its expensive processing (feature selection and evolutionary search) for each generated rule, while DTRM performs its expensive processing (tree construction) once per $\tuple{C_s, C_r, a}$ tuple and can quickly extract multiple rules from a tree.

\myparagraph{Experiments with Sample Policies}

DTRM and DTRM$^-$ spend most of the time in phase 1 to learn decision trees. The averaged running times spent on phase 2 are less than 1 second for EMR\_15 and project-mangagment\_5, and are less than 3 seconds for healthcare\_5 and university\_5.  DTRM and DTRM$^-$ are faster than FS-SEA* on all of these policies.  The average speedup is 2.17 for DTRM and 2.38 for DTRM$^-$.  DTRM has similar running time as DTRM$^-$ on the sample policies, since only a few negative features are generated when learning decision trees for these policies, and they are not useful and hence are removed in the ``{\em merge and simplify rules}'' phase, so the negative feature elimination step in DTRM has no work to do.
\myparagraph{Experiments with Case Study Policies}

For eWorkforce, the average speedup is 1.70 for DTRM and 1.96 for DTRM$^-$.  The negative feature elimination step in DTRM has little effect on the speedup, since the decision trees generated from the first phase do not contain many negative features. For e-document, the average speedup is 2.92 for DTRM and is 8.13 for DTRM$^-$, and the speedup for DTRM$^-$ increases with policy size.  The difference in speedup is larger for e-document, because more negative features are generated in phase 1, so the negative feature elimination step in DTRM takes longer.

DTRM and DTRM$^-$ have lower average speedups on sample policies and eWorkforce, compared with the other policies (discussed next), because these policies are simpler, allowing FS-SEA* to have relatively good running time on them.   In particular, FS-SEA* needs only one or a few iterations of feature selection and evolution to learn the rules for a given combination of subject type, resource type, and action, whereas for the more complicated policies, FS-SEA* typically needs more such iterations.

\myparagraph{Experiments with Synthetic Policies}

In experiments with the first family of synthetic policies, with $M=20$ rules and varying object model size, the average speedup is 12.31 for DTRM and 13.49 for DTRM$^-$.  For both DTRM and DTRM$^-$, the speedup generally increases with object model size; the 3\% dip from syn\_25\_20 to syn\_30\_20 is not statistically significant (it's less than the $\sigma$).



In experiments with the second family of synthetic policies, with object model size $N = 20$ and varying number of rules, the average speedup is 17.48 for DTRM and 20.76 for DTRM$^-$. The speedups of both DTRM and DTRM$^-$ significantly increase with the number of rules: for DTRM, speedup increases from 5.66 with 10 rules to 25.67 with 40 rules; for DTRM$^-$, speedup increases from 6.61 with 10 rules to 30.21 with 40 rules.



\subsection{Comparison with Iyer et al.'s Algorithm}
\label{sec:evaluation-results:compare-Iyer}

\begin{table*}[tbp]
\resizebox{\textwidth}{!}{%
\begin{tabular}{|l|l|l|l|l|l|l|l|l|l|l|l|l|}
\hline
\multicolumn{1}{|c|}{\multirow{2}{*}{Policy}} & \multicolumn{4}{c|}{Input Policies} & \multicolumn{3}{c|}{Avg. \# of Mined Rules} & \multicolumn{5}{c|}{Avg. Running Time (sec)} \\ \cline{2-13} 
\multicolumn{1}{|c|}{} & \#obj & \#field & \#FtVec & \#rules & \cite{iyer2019} & DTRM & DTRM$^-$ & \cite{iyer2019} & DTRM & SpdUp & DTRM$^-$ & SpdUp \\ \hline
eWorkforce\_10 & 354 & 530 & 8662 & 7 & 8 & 7 & 7 & 14 & 3 & 4.67 & 3 & 4.67 \\ \hline
eWorkforce\_15 & 505 & 751 & 20170 & 7 & 8 & 7 & 7 & 34 & 10 & 3.40 & 10 & 3.40 \\ \hline
eWorkforce\_20 & 601 & 897 & 29158 & 7 & 8 & 7 & 7 & 78 & 19 & 4.11 & 19 & 4.11 \\ \hline
eWorkforce\_25 & 755 & 1121 & 48253 & 7 & 8 & 7 & 7 & 101 & 42 & 2.40 & 41 & 2.46 \\ \hline
eWorkforce\_30 & 888 & 1304 & 67653 & 7 & 8.3 & 7 & 7 & 346 & 76 & 4.55 & 74 & 4.68 \\ \hline
\end{tabular}%
} \smallskip
\caption{Comparison of DTRM, DTRM$^-$ and Iyer et al.'s algorithm on the simplified eWorkforce\_10 dataset.  \#obj, \#field, \#FtVec and \#rules have the same meanings as in Figure \ref{fig:eval-method}. SpdUp is the speedup of DTRM and DTRM$^-$ relative to Iyer et al.'s algorithm. }
\label{tab:compare-with-Iyer}
\end{table*}

%

We compare DTRM and DTRM$^-$ with Iyer et al.'s ReBAC mining algorithm \cite{iyer2019} using modified versions of the eWorkforce datasets described in  Section \ref{sec:evaluation-results:compare-FS-SEA}.  We use Iyer et al.'s translation of a subset of the eWorkforce rules (used in experiments in \cite{iyer2019}) as a starting point, and update it retain more of the original ORAL2 rules.  We also modify the ORAL2 rules to exactly match (in  meaning and structure, not syntax) the translated rules.  We end up with 17 rules in each framework.
Note that the original eWorkforce rules cannot be used directly: they need to be simplified, because Iyer et al.'s framework in \cite{iyer2019} is less expressive than ORAL2.  In particular, we eliminate Boolean attributes, and set comparison operators other than equality.  We also simplify the object models in the eWorkforce\_10 dataset by eliminating fields and classes not used in the modified rules.  We implemented a translator that converts the simplified object models into Iyer et al.'s ``system graph'' representation.  This enables us to run their system on significantly larger system graphs than used in any of the experiments (with any policy, not just eWorkforce) in \cite{iyer2019}.  We then compare the results of running our algorithms and their implementation of their algorithms. 


When run on the modified eWorkforce\_10 dataset, their system does not finish in a reasonable time (we used a timeout of 30+ minutes, since DTRM and DTRM$^-$ take less than a minute for this dataset) for some object models, and it returns errors, such as ``MemoryError'' and ``IndexError: pop from empty list'', for others.  We reported these issues to Iyer et al.  Until they provide a fix, we circumvented these issues by removing the rules that trigger these issues, and removing parts of the object models unused by the remaining rules, until their system ran successfully for the remaining rules and at least one of the simplified object models for each object model size.  In the end, we removed 10 rules that their system has trouble with, leaving 7 rules.  The majority of the problematic rules are syntactically more complicated than the remaining ones.  Specifically, 8 out of 10 of the problematic rules contain more than two atomic conditions/constraints (in ORAL2) or relationship patterns (in \cite{iyer2019}'s policy language).  In contrast, most (specifically, 5 out of 7) of the remaining rules contain only one atomic condition/constraint or relationship pattern (the other two remaining rules contain 3 atomic conditions/constraints). 
Results of these experiments are reported in Table \ref{tab:compare-with-Iyer}.  We set the path length limits for DTRM and DTRM$^-$ to smaller values suitable for these simplified policies: $\mcse = 5$, $\mspl=2$, $\mrpl=1$, $\sped=0$, $\rped=0$, and $\mtpl=4$.  Even using these 7 remaining rules and significantly simplified object models, their system does not finish in a reasonable time (30 minutes) for some of the 5 object models for each policy size.  Although we do not know for certain whether this is due to inefficiency of their algorithm or bugs in their implementation, we make the more generous assumption (i.e., assume the latter) and therefore omit those object models from the reported results.  Consequently, the results in Table \ref{tab:compare-with-Iyer} are averages over 4 object models for eWorkforce\_10, 1 for eWorkforce\_15, 1 for eWorkforce\_20, 2 for eWorkforce\_25, and 3 for eWorkforce\_30. 


All three algorithms mine policies that grant the same authorizations as the input policies.  For DTRM, the mined policies are identical to the input policies. For DTRM$^-$, the mined policies are almost identical to the input policies: the only difference is replacement of the condition tenant.id = PP in one input rule with the negative condition tenant.id $\ne$ Telco, which is equivalent in context of these simplified object models.  For Iyer et al.'s algorithm, the mined policy contain one more rule than the original policy (8 instead of 7) for all object models, except it contains two more rules for one object model of eWorkforce\_30, because their algorithm fails to mine some of the desired relationship patterns, generating instead multiple rules containing longer relationship patterns.
We do not report WSC for these experiments, because the algorithms use different policy languages, and WSC is language-dependent.

DTRM and DTRM$^-$ are faster than Iyer et al.'s algorithm for all policies. Averaged over all policies, DTRM is 3.83 times faster, and DTRM$^-$ is 3.86 times faster.  DTRM and DTRM$^-$ have very similar running times in these experiments, because very few negative features appear in the rules extracted from the decision trees.


\section{Future Work}
\label{sec:conclusion}


Directions for future work include: extending our algorithms to handle incompleteness and noise in the ACLs, perhaps using decision tree pruning methods, which are designed to avoid overfitting; extending our algorithms to identify errors in attribute values, and possibly suggest corrections; 
and developing incremental algorithms that efficiently handle updates to the object model or authorizations.


\articleonly{\paragraph{Acknowledgements.} \thanksText }


\svonly{
\begin{acknowledgements}
\thanksText
\end{acknowledgements}}


%
\ieeeonly{\bibliographystyle{IEEEtran}}\acmonly{\bibliographystyle{ACM-Reference-Format}}\lncsonly{\bibliographystyle{splncs04}}\articleonly{\bibliographystyle{alpha}}\svonly{\bibliographystyle{plain}}
\bibliography{references}

 \end{document}